\documentstyle[pra,aps,preprint]{revtex}
\begin{document}

\preprint{SNUTP 98-067}
\draft
\title{
Heisenberg picture approach to the invariants and the exact 
quantum motions for coupled parametric oscillators
}

\author{
Jeong-Young Ji\footnote{Electronic address: jyji@phya.snu.ac.kr} 
and
Jongbae Hong\footnote{Electronic address: jbhong@phyc.snu.ac.kr}
}
\address{Department of Physics Education, Seoul National University, 
Seoul 151-742, Korea
}

\maketitle

\begin{abstract}
For $N$-coupled generalized time-dependent oscillators, primary 
invariants and a generalized invariant are found in terms of classical 
solutions. Exact quantum motions satisfying the Heisenberg equation of 
motion are also found. For number states and coherent states of the 
generalized invariant, the uncertainties in positions and momenta are 
obtained.
\end{abstract}


In the time-dependent coupled oscillator system, the invariant method is 
powerful in analyzing the quantum mechanical behaviors. The 
Lewis-Riesenfeld (LR) invariant~\cite{Lewis,LR69} has been derived by 
various methods such as time-dependent canonical 
transformations~\cite{Leach77,Leach79}, 
Noether theorem~\cite{Lutzky78},
and Ermakov's technique~\cite{RayR79}. While the LR invariant is 
quadratic in position and momentum operators, the primary invariant found 
in Ref.~\cite{MalkM70} is linear in the operators. 

The primary invariant is so simple in structure that it may be useful in 
studying time-dependent coupled oscillators. Recently, Ermakov-Lewis 
invariant has been constructed using amplitude-phase decomposition in a 
coordinate-coordinate coupled systems~\cite{ThylweK98}. For the most 
general form of coupled oscillators which includes any couplings of 
coordinates and momenta, the LR-type invariant was found using the 
canonical transformation~\cite{Leach77} and the first-order invariant was 
constructed using the Noether theorem~\cite{CastLM94}.

In the Heisenberg picture, the quantum motions of position and momentum 
in a single oscillator system have been found in Ref.~\cite{JiKK95}, where 
the LR invariant exhibits the time-independency explicitly. In this 
Letter, we extend the previous work~\cite{JiKK95} to the coupled 
oscillators. Since the coupled parametric oscillator system studied here 
is the most general form, our results will be applied to the studies in 
quantum optics as well as in atomic and molecular physics. We construct 
primary invariant and LR invariant in terms of classical solutions and 
find the time-evolutions of position and momentum operators  which are the 
solutions of the Heisenberg equations. 

Let us consider a general oscillator type Hamiltonian
\begin{equation}
H(t) = A_{\mu \nu} (t) z^{\mu} z^{\nu} + B_{\mu} (t) z^{\mu} + C(t)
\label{H(t)}
\end{equation}
where the matrix $A_{\mu \nu} $ is real and symmetric.
As a unified notation for the coordinates in phase space, we use 
$\{ z^{\mu} \} $ : $q_{i} = z^{i} , ~
p_{i} = z^{N+i} $. Here all Greek indices $\alpha, ~\beta, ... $ range 
from $1$ to $2N$ and Latin indices $i, ~j,... $ range from $1$ to $N$. The 
symplectic matrix $[ \epsilon^{\mu \nu} ] $ and its inverse matrix 
$[ \epsilon_{\mu \nu} ] $ are defined by
\begin{equation}
\epsilon \equiv [ \epsilon^{\mu \nu} ] = 
{ \left(\begin{array} {cc} ~ 0 & 1_{N}  \\ -1_{N} & 0 
 \end{array}\right)}, ~
[ \epsilon_{\mu \nu} ] = 
{ \left(\begin{array} {cc} 0 & - 1_{N}  \\ 1_{N} & ~ 0 \end{array}\right)},
\label{sym}
\end{equation}
where $1_{N} $ is an $N \times N $ identity matrix.

We look for the primary invariant of the form
\begin{equation}
b = v_{\nu} (t) z^{\nu} (t) + u (t)
\label{b:def}
\end{equation}
which satisfies the invariant equation
\begin{equation}
\frac{{\partial }}{\partial t} b (t) - i [ b (t) , H(t) ] = 0 .
\end{equation}
From this invariant equation and the commutation relation 
$[ z^{\mu} , z^{\nu} ] = i \epsilon^{\mu \nu} $,  we have a system of the 
first order differential equations for $v_{\nu} $ and $u $:
\begin{eqnarray}
\dot{v}_{\nu} + 2 v_{\sigma} \epsilon^{\sigma \rho} A_{\rho \nu} = 0 , 
\label{vde}  \\
\dot{u} + v_{\nu} \epsilon^{\nu \sigma} B_{\sigma} = 0 .
\label{ude}
\end{eqnarray}
Note that Eq.~(\ref{vde}) is identical with the homogeneous part of the 
classical equation of motion (Hamilton's equation) for Eq.~(\ref{H(t)}):
\begin{equation}
\dot{z}_{{\rm cl}, \nu} = 
- 2 z_{{\rm cl}, \sigma} \epsilon^{\sigma \rho} A_{\rho \nu} + B_{\nu}
\end{equation}
with $z_{{\rm cl}, \nu} = \epsilon_{\nu \rho} z_{\rm cl}^{\rho} . $
For Eq.~(\ref{ude}), the solution of $u$ is easily found by the direct 
integration
\begin{equation}
u(t) = u( 0 )  - \int_{0}^{t} v_{\nu} (s) \epsilon^{\nu \sigma} B_{\sigma} 
 (s) ds .
\label{solu}
\end{equation}

If we represent the solution of Eq.~(\ref{vde}) as a complex row vector, 
there exist $2N$ linearly independent solutions which we label 
${\bf v}^{(1)} (t), ... ,{\bf v}^{(2n)} (t). $ Combining these solutions 
together we define the solution matrix
\begin{equation}
V=[ v_{~\nu}^\mu ] = 
{ \left(\begin{array} {c} {\bf v}^{(1)}  \\ \vdots  \\ {\bf v}^{(2n)} \end{array}\right)}
\end{equation}
which obeys
\begin{equation}
\dot{V} = 
- 2 V \epsilon A .
\label{vdem} 
\end{equation}

This solution matrix is not determined uniquely because the linear 
combination of solutions is also a solution. In other words, if $V$ is a 
solution matrix, so is $CV$ with a nonsingular constant matrix $C$. We 
choose the solution matrix of the form:
\begin{equation}
V = i { \left(\begin{array} {cc}
- \pi^{*} & \phi^{*}  \\ 
 \pi & - \phi \end{array}\right)}
\label{V} 
\end{equation}
satisfying the following initial conditions 
with arbitrary parameters $\omega_{i}  $,
\begin{equation}
\phi_{ij} (0) = \frac{1}{\sqrt{2 \omega_{i}}} \delta_{ij} , ~
\pi_{ij} (0) = - i \sqrt{\frac{\omega_{i}}{2} } \delta_{ij} ,
\end{equation}
where $\phi = [ \phi_{ij} ] $ and $\pi = [ \pi_{ij} ] $ are $N \times N$ 
matrix.
Then we have the following $2N$-primary invariants
\begin{equation}
b^{\mu} = v_{~\nu}^\mu (t) z^{\nu} (t) + u^{\mu} (t)
\label{bs}
\end{equation}
which satisfy (i) $b^{i} = b_{i} , ~b^{N+i} = b_{i}^{\dagger} $ 
(ii) $[ b^{\mu} , b^{\nu} ] 
= v_{~\alpha}^\mu  v_{~\beta}^\nu i \epsilon^{\alpha \beta} 
= \epsilon^{\mu \nu} . $
These conditions (i) and (ii) mean that we can interpret 
$b^{i} (b^{i \dagger}) $ as the annihilation (creation) operator. 
Inversely, position and momentum operators are given by

\begin{eqnarray}
q_{i} (t) &=& \phi_{ij} b_{j} - \phi_{ij} u_{j} + {\rm h.c.}  \\
p_{i} (t) &=& \pi_{ij} b_{j} - \pi_{ij} u_{j} + {\rm h.c.} 
\label{qp:bb}
\end{eqnarray}
where we have used
\begin{equation}
V^{-1} = i \epsilon V^{T} \epsilon^{T} = 
{ \left(\begin{array} {cc} \phi & \phi^{*}  \\
\pi & \pi^{*} \end{array}\right)} .
\end{equation}

With the first-order invariants, the LR type invariant is constructed as
\begin{equation}
I = \sum_{i}^{N} \omega_{i} \left( b_{i}^{\dagger} b_{i} + \frac{1}{2} 
 \right) .
\label{I}
\end{equation}
The eigenstates of the invariant (\ref{I}) are the number states
\begin{equation}
\left| \bf n \right> \equiv \left| n_{1} , ..., n_{N} \right> = 
\prod_{i} \frac{{ b_{i}^{\dagger n_{i} } }}{\sqrt{n_{i} !}} 
\left| \bf 0 \right>
\label{nstate}
\end{equation}
where the state $\left| \bf 0 \right> $ is defined, as usual, by
\begin{equation}
b_{i} \left| { \bf 0 } \right> = 0, ~ {\rm for} ~ i = 1, ..., N .
\end{equation}
Further, we define the coherent state as
\begin{equation}
\left| \mbox{\boldmath$ \alpha$} \right> = 
\prod_{i=1}^N
e^{- | \alpha_{i} |^{2} / 2} \sum_{k_{i} =0}^\infty 
\frac{\alpha_{i}^{k_{i}}}{\sqrt{k_{i} !}}
\left| k_{i} \right> 
\label{cstate}
\end{equation}
which satisfies
\begin{equation}
b_{i} \left| \mbox{\boldmath$ \alpha$} \right> = \alpha_{i} \left| 
 \mbox{\boldmath$ \alpha$} \right> .
\end{equation}

In the Heisenberg picture the time evolution of the system is described 
by the time evolution of the quantum operators. By equating the invariants 
(\ref{bs}) in two different times:
\begin{equation}
v_{~\nu}^\mu (t) z^{\nu} (t) + u^{\mu} (t) =
v_{~\nu}^\mu (0) z^{\nu} (0) + u^{\mu} (0) 
\label{btb0}
\end{equation}
we deduce the quantum evolution of the Heisenberg operators
\begin{equation}
{\bf z} (t) = V^{-1} (t) V(0) {\bf z} (0) 
- V^{-1} (t) [ {\bf u} (t) - {\bf u} (0) ] .
\label{zt}
\end{equation}
By direct differentiation, it is easily checked that $z^{\mu} (t) $ 
satisfies the Heisenberg equation of motion 
$i \frac{d}{dt} z^{\mu} = [z^{\mu} , H ] $. In the explicit form of 
positions and momenta, we obtain
\begin{eqnarray}
q_{i} (t) &=& 
-i \phi_{ij} (t) \pi_{jk}^* (0) q_{k} (0)
+i \phi_{ij} (t) \phi_{jk}^* (0) p_{k} (0)
- \phi_{ij} (t) [ u_{j} (t) - u_{j} (0) ] + {\rm h.c.},
 \\
p_{i} (t) &=& 
- i \pi_{ij} (t) \pi_{jk}^* (0) q_{k} (0)
+ i \pi_{ij} (t) \phi_{jk}^* (0) p_{k} (0)
- \pi_{ij} (t) [ u_{j} (t) - u_{j} (0) ] + {\rm h.c.}.
\label{pqt}
\end{eqnarray}
These results can be easily obtained from (\ref{qp:bb}) replacing $b_{i}  $ 
by (\ref{bs}) at $t = 0  $.

Now we examine the quantum properties of the eigenstate and the coherent 
state. The variations in position and momentum are given by, for the 
number state (\ref{nstate}), 
\begin{eqnarray}
\left< {\bf n} \right| 
( \Delta q_{i} )^{2} (t) 
\left| {\bf n} \right>
&=& \sum_{j} (2n_{j} + 1) |\phi_{ij}|^{2} ,
 \\
\left< {\bf n} \right| 
( \Delta p_{i} )^{2} (t) 
\left| {\bf n} \right>
&=& \sum_{j} (2n_{j} + 1) |\pi_{ij}|^{2} ,
\label{dqdpn}
\end{eqnarray}
and for the coherent state (\ref{cstate})
\begin{eqnarray}
\left< \mbox{\boldmath$ \alpha$} \right| 
( \Delta q_{i} )^{2} (t) 
\left| \mbox{\boldmath$ \alpha$} \right>
&=& \sum_{j} |\phi_{ij}|^{2} ,
 \\
\left< \mbox{\boldmath$ \alpha$} \right| 
( \Delta p_{i} )^{2} (t) 
\left| \mbox{\boldmath$ \alpha$} \right>
&=& \sum_{j} |\pi_{ij}|^{2} .
\label{dqdpa}
\end{eqnarray}
Furthermore the expectation values of the position and the momentum for 
the coherent state is 
\begin{eqnarray}
\left< \mbox{\boldmath$ \alpha$} \right| 
q_{i} (t) 
\left| \mbox{\boldmath$ \alpha$} \right>
&=& 
\phi_{ij} \alpha_{j} - \phi_{ij} u_{j} + {\rm c.c. } ,
 \\
\left< \mbox{\boldmath$ \alpha$} \right| 
p_{i} (t) 
\left| \mbox{\boldmath$ \alpha$} \right>
&=& 
\pi_{ij} \alpha_{j} - \pi_{ij} u_{j} + {\rm c.c. } ,
\label{aqpt}
\end{eqnarray}
which are the same as those of the classical motion.

Finally we suggest that our formalism can be applied to the following 
type of Hamiltonian
\begin{equation}
H(t) = { \cal A }_{\mu \nu} a^{\mu} a^{\nu} + {\cal B}_\mu a^{\mu} + C
\end{equation}
where
\begin{equation}
{\cal A}_{\mu \nu} = 
A_{\rho \sigma} \Lambda_{~\mu}^\rho \Lambda_{~\nu}^\sigma , ~
{\cal B}_\mu = B_{\rho} \Lambda_{~\mu}^\rho .
\end{equation}
Here we have introduced the creation and annihilation operators defined 
with an arbitrary parameter $\lambda $ by
\begin{eqnarray}
q_{i} &=& \sqrt{ \frac{1}{ 2 \lambda} } ( a_{i} + a_{i}^{\dagger} ),  \\
p_{i} &=& \frac{1}{i} \sqrt{ \frac{\lambda}{2} } ( a_{i} - 
 a_{i}^{\dagger} ) ,
\label{qp:aa}
\end{eqnarray}
or in the unified notation $(a_{i} = a^{i} , ~ a_{i}^{\dagger} = a^{N+i} ) $
\begin{equation}
z^{\mu} = \Lambda_{~\nu}^\mu a^{\nu}
\end{equation}
with
\begin{equation}
\Lambda = { \left(\begin{array} {cc}
\sqrt{ \frac{1}{2 \lambda} } 1_{N} & 
\sqrt{ \frac{1}{2 \lambda} } 1_{N}  \\
\frac{1}{i} \sqrt{ \frac{\lambda}{2} } 1_{N} & 
- \frac{1}{i} \sqrt{ \frac{\lambda}{2} }  1_{N} 
\end{array}\right)} .
\end{equation}

In summary, we find the primary(first-) and second-order(LR-type) 
invariants of the $N$-coupled generalized time-dependent oscillators. 
Using the primary invariants, we find the exact quantum motions for the 
position and the momentum operators satisfying the Heisenberg equation of 
motion. The primary invariants give very simple method to find the quantum 
motions. We also studied the quantum properties of the eigenstates and the 
coherent states for the LR-type invariant. As in the case of the single 
parametric oscillator~\cite{JiKK95,KimLJK96,LeeKJ97}, the classical 
solutions give all the descriptions of the corresponding quantum system.

This work was supported by the Korean Science and Engineering Foundation 
and the Center for Theoretical Physics (SNU).

\end{document}